\documentclass[a4paper,11pt]{article}
\usepackage[a4paper,top=3cm,bottom=3cm,left=2cm,right=2cm,marginparwidth=1.75cm]{geometry}
\usepackage[toc,page]{appendix}
\usepackage[english]{babel}
\usepackage{polski}
\usepackage[utf8]{inputenc}
\usepackage{amsmath}
\usepackage{amssymb}
\usepackage{lmodern}
\usepackage{color}
\usepackage{nccmath}
\usepackage{graphicx}
\usepackage{wrapfig}
\usepackage{fancyhdr}
\usepackage{enumerate}
\usepackage{geometry}
\numberwithin{equation}{section}
\usepackage{hyperref}
\usepackage{braket}
\usepackage{float}
\usepackage{dsfont}
\usepackage{indentfirst}
\DeclareMathOperator*{\Tr}{Tr}

\definecolor{darkmagenta}{rgb}{0.55, 0.0, 0.55}

\definecolor{darkolivegreen}{rgb}{0.33, 0.42, 0.18}
\newcommand{\OG}[1]{\textcolor{darkolivegreen}{#1}}

\begin{document}
\begin{titlepage}
    \begin{center}
        \vspace*{1.5cm}
  \Large
      University of Wrocław\\
      Department of Physics and Astronomy\\
      Institute of Theoretical Physics\\
        \mbox{}
        \\
  \Large
       Maciej Kowalczyk\\
        \Huge
        \textbf{Consequences of
regularization ambiguities in Loop Quantum Cosmology}\\
        \vspace{0.5cm}
        
 \Large
       Konsekwencje niejednoznacznej regularyzacji w Pętlowej Kosmologii Kwantowej

        \vspace{2.5cm}
    \end{center}
\begin{flushright}
Supervisor of master thesis\\
dr hab. Tomasz Pawłowski
\end{flushright}
\begin{center}
\vspace*{4.5cm}
Wrocław 2021
\end{center}
\end{titlepage}
\newpage
\thispagestyle{empty}
\begin{abstract}
\vspace{0.5cm}
Ambiguities of the so-called Thiemann regularization in Loop Quantum Cosmology lead to freedom in how to construct a particular quantization prescription. So far three distinct examples of such have been proposed in the literature. For two of them, detailed analysis has been already performed in the literature. In this thesis, the methodology developed for these is applied to study in detail the third one, which will be referred to as mLQC-II. In particular, the mathematical properties of the operator of the quantum version of full Hamiltonian constraint are examined. These properties indicate that the evolution of the system is uniquely determined. Furthermore, an investigation of dynamics is performed by finding the expectation value of the volume as a function of the scalar field and constants of motion. This result is next compared to the trajectories predicted with the so-called effective dynamics. Finally, the main properties of studied prescription are compared against those of other (already investigated) ones. \vspace{2cm}

Niejednoznaczności wynikające z tak zwanej regularyzacji Thiemana w Pętlowej Kosmologii Kwantowej prowadzą do swobody w procedurze kwantowania. Do tej pory wyróżniono trzy konkretne wybory, z których dwa zostały już poddane szczegółowej analizie. W niniejszej dysertacji użyjemy wypracowanej już metodologii w szczegółowej analizie trzeciego wyboru, do którego będziemy odnosić się jako mLQC-II. W szczególności analizujemy własności operatora kwantowego wiązu hamiltonowskiego, które wskazują na jednoznacznie zdeterminowaną ewolucję systemu. Ponadto badanie dynamiki przeprowadzono poprzez obliczenie wartości oczekiwanej objętości jako funkcji zależnej od pola skalarnego oraz stałe ruchu. Wynik ten jest następnie porównywany z tymi otrzymanymi już w literaturze dla innych recept regularyzacji. \vspace{2cm}

~~\\
~~\\~~\\
\indent Słowa kluczowe: Ogólna Teoria Względności, Kwantyzacja Polimerowa, Kosmologia Kwantowa, Pętlowa Kosmologia Kwantowa
\end{abstract}
\newpage
\thispagestyle{empty}
\mbox{}
\tableofcontents
\newpage
\setcounter{page}{2}
\section{Introduction} 
The variety of gathered knowledge forces us to organize it into fields in order not to get lost in the richness of information. The one dedicated to finding the most fundamental rules governing the observed reality is Physics. When trying to comprehend the fundamental interactions within matter, the most successful ,,component'' up to date is Quantum Physics: initialized by a theory that started growing in the last century - Quantum Mechanic and subsequently expanded into Quantum Field Theory and in particular the Standard Model neatly systematizing the richness of forms of observable matter (on the level of particle physics) and known non-gravitational interactions. On the other hand, if gravity influence on the matter is the main course of consideration, then General Relativity (provided that quantum effects are abandoned) delivers, with great precision, answers about the Universe. Descriptions of nature provided by QM and GR have been confirmed by numerous experiments in both cases but they arise from principles that are in partially in contradiction to each other. In the quantum approach, we operate only with probabilities  (thus expectation values) rather than exact (classical) trajectories like in the theory of relativity. Departure becomes even more explicit if we take look at spacetime: in QM it is fixed (mainly Minkowski) space which serves as a background to describe interaction - one can still consider Quantum Field Theory in the curve spacetime. While GR metric evolves with matter and cannot be simply fixed. So it is impossible to build one description of nature by just simply adding these two theories. In addition, we still encounter several open problems that are not covered alone by any of them. For example, observation shows that most of the matter in our universe cannot be described by the SM. The other one is the absence of a theory that describes the quantum behaviour of gravity along with experimental confirmation. But it doesn't mean that any attempt forward such theory wasn't made. For example, if one chooses as a starting point particle physics one ends up with an elegant theory with extra dimensions - the String Theory. Because of its mathematical elegance and the fact that this theory combines efforts in the quantization of gravity with hopes of unification, it was the most examined course at the end of the previous century \cite{Rovelli:1997qj}. \\ 
\indent But nature not necessarily has to been ,,enclosed'' by a single theory. One of alternatives branch of propositions for String Theory are attempts at a non-perturbative and background independent quantization of General Relativity. Examples of them are quantization of Geometrodynamics which to this day is not finished due to technical obstacles, or one that performs quantization of GR in loop variables - Loop Quantum Gravity. To point out the beginning of this branch, it is sufficient to go back to the XX century, exactly to the 30s to the work of Matvei Bronstein. He was a young soviet researcher who encouraged by the debate between Bohr, Rosenfeld and Landau, Peierls about the application of the uncertainty principle to the electromagnetic field, carried out similar consideration with respect to the gravitational field \cite{bronestein, Rovelli:2014ssa}. What he found out is that when GR and QM play a role, it is not possible to localize anything with a precision better than the Planck length ($10^{-35}$m). It means that at some scales to describe the world we need to use both quantum and gravitational properties. This claim is supported by a conclusion of similar consideration made by Hawking who applied statistical quantum mechanics to black holes. What was obtained is a form of black body radiation which cause black hole evaporation. It confirmed that black holes obey the second law of thermodynamic and are encode (alongside information about Planck length) in the Bekenstein-Hawking entropy formula. Several years after this revelation, namely in 1986 General Relativity has been re-formulated in connection variables. This moment is usually identified as a starting point in the development of LQG. In the following months, the solution of the Wheeler-DeWitt equation was found in the form of Wilson Loops. From that moment, LQG started to gain shape with the establishment of loop representation in 1987 and the introduction of the scalar product in 1994. In the same year, one of the most important results of LQG was shown - discreteness of area and volume eigenvalues \cite{Rovelli:1994ge}. As it is a relatively young theory it is still under vigorous development. Its flaw is, as in the case of String theory, that still no experiment that could confirm or discard it has been performed \cite{Rovelli:1997yv, Rovelli:1997qj}. But during the development of LQG subfield by the name Loop Quantum Cosmology emerge which is an application of methods of LQG to a cosmological setting. It was first introduced in 2002 in the work of Martin Bojwald \cite{Bojowald:2002gz} and in the next year, its mathematical structure has been discussed in detail in \cite{Ashtekar:2003hd}. In 2006 first solid calculation of dynamics for homogeneous isotropic models has been performed by Ashtekar, Pawłowski and Singh in \cite{Ashtekar:2006uz, Ashtekar:2006wn, Ashtekar:2006es} and in the following years to anisotropic systems (for example to Bianchi I spacetime in the vacuum \cite{MartinBenito:2008wx}). Some focus as well has been directed to inhomogeneous models \cite{MartinBenito:2008ej}. Results obtained in these and many other papers confirms that this application can predict physically interesting scenarios.

Because LQC, as well as LQG (and GR for that matter), is a constraint theory, in order to probe the dynamics within it one has to build and study Hamiltonian constraint. The matter part of the constraint is not problematic but the construction of the gravity part is not straightforward \cite{Ashtekar:2011ni, Ashtekar:2008zu}. To do so one first promote elementary variables (in LQC holonomy and densitized triads) to the operators and then by a procedure known as Thiemann regularization build the constraint, which procedure is not unique. This ambiguity in consequence can provide different models. One originally presented in \cite{Ashtekar:2006uz} is referred to as a mainstream LQC. It utilizes the fact that in the spatially flat Friedmann Lemaitre Robertson Walker spacetime after symmetry reduction Euclidean and Lorentzian terms of gravity part of hamiltonian become proportional.  Since that, the Lorentzian term can be rewritten in the form of the Euclidean one on the classical level and then merge to the Euclidean part. The next step is to quantize this term. The evolution of this system leads to a single quantum bounce at near-Planckian energy densities, connecting two epochs of a large semiclassical Universe. But at it was first shown in \cite{Yang:2009fp} when Euclidean and Lorentzian parts are treated separately, two modifications of LQC can be obtained. One referred, in the literature, as mLQC-I is obtained by quantizing Euclidean and Lorentzian terms independently, while one referred as mLQC-II additionally using fact that extrinsic curvature in k=0 FLRW is proportional to Ashtekar-Barbero connection. mLQC-I has been already well examined in the literature (eg. \cite{Assanioussi:2019iye}) with various novelties with respect to LQC (for example rather than one bounce two bounces occur with two emergent DeSitter epochs between them) so the goal of this thesis is to examine mLQC-II to check properties of the operator of full Hamiltonian constraint and to obtain expectation value of volume operator and it semiclassical approximation.

This thesis is organized as follows. In the following section, basic information about formalism needed further will be introduced via a very general outline of LQG and through examples of LQC and mLQC-I. The content of the main section will be fully dedicated to the mLQC-II - to building Hamiltonian constraint, to finding the evolution operator $\theta$ and scalar product in the different representations and most important to finding expectation values of volume operator. The last chapter will serve to discuss and compare obtained results. 

\section{Formalism}
Let us start with introducing briefly the formalism applied in this
dissertation. First, in section \ref{intro} we will briefly
present the main ideas and components of Loop Quantum Gravity as the
most general but not yet complete framework. Next, in
sec.~\ref{lqc-intro} we will introduce the core elements of Loop
Quantum Cosmology as the symmetry reduced version of LQG, discussing how
the ideas behind LQG are realised in the context of cosmological models.
Finally, a version of LQC born of applying directly the regularization
prescription introduced by Thiemann in LQG will be presented in
sec.~\ref{mlqc}.

\subsection{Loop Quantum Gravity} \label{intro}

Since Loop Quantum Gravity implements strictly the main principles of General Relativity it
treats spacetime itself and the gravitational field as the aspects of the same physical
entity. Dividing the spacetime metric into a fixed background metric and fluctuating quantum field is
thus forbidden. So it is a non-perturbative approach. Moreover, it can be split into two branches -
covariant (the so-called spin foam models) and the canonical one.\\
\indent Here we focus on the latter, where one performs foliation of the space-time described by a 4-dimensional manifold with some metric into a set of 3-dimensional space-like surfaces $\Sigma_t$ ($t$ denotes the global time function) \cite{Kiefer:2007ria, Rovelli:2014ssa}. The spacetime $4$-metric spits then onto spatial $3$-metric (on $\Sigma_t$), a lapse function and a shift vector. In such case, elementary variables are taken to be 3-metric and its conjugate momentum. But 3-metric can be expressed in terms of orthonormal co-triad. Thus as a set of canonically conjugate variables,  one can chooses a Sen-Ashtekar-Barbero variables: $su(2)$ valued connection $A_a^i$ (indices $a=1,2,3$ are spatial indices and $i=1,2,3$ enumerating vectors) and densitized triads $E_i^a$. Now, the starting point - like in ADM formulation - is the Einstein-Hilbert action written in terms of
new variables and with appropriate boundary term (Holst term). Its canonical splitting and Legendre transform leads to a standard set of
constraints of ADM theory: Hamiltonian and diffeomorphism one. However in this Hamiltonian formulation emerge additional constraint, which ensures that the
state belonging in the system is invariant under gauge transformation - it is called Gauss constraint.  Since it is a system with constraint, its quantization
requires special care. Here the systematic method known as the Dirac program
is used.\\
\indent At first quantization is performed only on the kinematical level - ignoring the constraints. But presented variables are not however direct candidates for quantization. Instead, one takes the holonomies of connection along curves and fluxes of $E_i^a$ across surfaces.  They form a (so called) holonomy
flux $*$-algebra, which allows to use the machinery of the Gellfand-Neumark-Segal method to build a quantum theory. It produces non-separable kinematical Hilbert space spanned by spin-networks states represented by cylindrical functions \cite{Husain:2011tk} living on the piecewise-analytic graphs (embedded in 3D manifold). In addition state living on an empty graph is distinguished as a vacuum.

The next step is to implement the constraints and for each of them, a different method is used. Gauss constraint implies certain restrictions
on the spin labels of edges intersecting particular vertex and allows to
distinguish a sub-basis of the chosen basis of the original kinematical
Hilbert space. The diffeomorphisms are implemented as finite
transformations not via infinitesimal generators (the constraint). The
diffeomorphism invariant states are identified by the procedure of the so-called group averaging \cite{Ashtekar:1995zh} - are literary averages (integrals) over the action of all the spatial diffeomorphisms. At this subspace manifest one of the most important results of LQG -
the discreteness of the geometry. Now all the properties of continuous geometry (for example volumes or angles) are represented by quantum operators. In particular, most relevant for LQG is the area one which features purely discrete spectra \cite{Thiemann:2007zz, Rovelli:1994ge}. Its minimum nonzero eigenvalue is $A_1=2\sqrt{3}\pi \gamma l_p^2$, where: $l_p$ is a Planck length and $\gamma$ is the Barbero-Immirzi parameter that in these paper will be treated (as suggested by the black hole entropy calculations \cite{Ashtekar:2004eh,Domagala:2004jt,Meissner:2004ju}) as a real number close to $0,2375$ (in first consideration of Ashtekar it was a pure imaginary number). At this point, one needs to be aware that neither of the above operators is physical observable. 

Finally, the Hamiltonian constraint is implemented by promoting the classical constraint to a quantum operator. It is done by reformulating it - through a procedure called Thiemann regularization - in terms of elementary variables. What is important here is that such an outcome is not unique.  The physical Hilbert space will then be identified as a kernel of that operator which is well defined mathematically, but due to the complicated nature of the operator extremely difficult technically. But one can reformulate the given theory as a free
theory by introducing matter fields - which serves as an evolution parameter - coupled to gravity \cite{Domagala:2010bm,Husain:2011tk,Giesel:2012rb}.  Thanks to that the Hamiltonian constraint operator becomes an evolution equation with the
gravitational part becoming a true Hamiltonian. According to the Dirac program, it is an appropriate time to construct physical observables. As such one uses the so-called partial observables \cite{Rovelli:2001bz, Dittrich:2004cb} measuring relevant quantities in relation to other ones. For example, if the choice of matter field allows identifying kinematical diffeomorphism invariant Hilbert space as a physical one, mentioned earlier operators can become partial observables measuring geometrical quantities with respect to the clock field.

It is worth pointing out, that consideration of above-described the theory is the gravitational field itself. Unfortunately, so far the only examples for which the quantum dynamics has been evaluated correspond to textbook examples of simplest spin networks \cite{Zhang:2019dgi}, not
corresponding to any relevant physical system. But one can use the above tools and
steps to focus on the entire universe as a quantum system without an
external observer. Such an approach simplify consideration and can provide
useful information -  for example in the regime of critical energy.

\subsection{Loop Quantum Cosmology}\label{lqc-intro}

Despite being called the symmetric sector of LQG, this approach is not so in the strict sense. It is not derived from full theory. Instead, it is an independent simplified approach constructed by applying the quantization program of LQG to cosmological (highly) symmetric models in a step-by-step basis. Here we are going to consider it in its simplest setting: applied to a model of a flat FLRW universe, which  metric in polar coordinates has the form: 
 \begin{equation}
 ds^2=dt^2-a^2(t)\Big(dr^2+r^2d\Omega^2\Big)
 \end{equation} 
In particular, our calculations will be performed for such a universe admitting the massless scalar field.

At this point we have the first departure from the diffeomorphism-invariant formalism of LQG: because homogeneous spacetimes admit additional distinguished structure, we partially fix the spatial gauge \cite{Ashtekar:2004eh}. In particular as the foliation by spatial surfaces, we choose the one distinguished symmetry - foliation by homogeneous surfaces. Furthermore, the isotropy allows to partially fix the gauge when selecting the orthonormal triad, in particular allowing to set the Ashtekar variables defined in LQG to take the form:
\begin{align}
A_a^i~&=~cV_0^{-\frac{1}{3}}\delta^i_a & E_i^a~&=~pV_0^{-\frac{2}{3}}\delta^a_i \ .
\end{align}
where $c$ and $p$ are constant on the spatial slice. The coefficient $V_0$ is a volume of the fiducial cell which define as the finite region of the Universe that remains constant in comoving coordinates. Choosing such a cell is necessary as all the momenta calculated by integrals of densities over the whole spatial slice would be infinite due to the homogeneity of spacetime. It means that all quantities evaluated in the process of dynamical evolution will refer only to that finite region.

The Poisson Bracket of connection and densities triad \eqref{A.1.5} becomes:
\begin{align}
\{c,p\}=\frac{\kappa\gamma}{3}
\end{align}
where $\kappa$ is constant equal to $8\pi G$. 

As in LQG, we proceed with choosing as our basic variables holonomies and fluxes. Here however the symmetry isotropy allows us to greatly simplify the holonomy flux algebra we work with. Instead of the full one, we can choose only the holonomies along the straight edges - in the direction $k$ - in $\Sigma_t$: 
\begin{equation}
h_k^{(\lambda)}=\cos\frac{\lambda c}{2} \mathds{1} +2\sin\frac{\lambda c}{2}\tau_k 
\end{equation} 
where $\lambda$ is any real number and $\tau_k$ are Pauli matrices (multiplied by $-\frac{i}{2}$) and the fluxes across unit square surfaces here also denoted as $p$ \cite{Ashtekar:2003hd}.

Having selected our basic variables we now proceed with the Dirac quantization program. In the first step (kinematic) we apply the GNS method to our reduced holonomy algebra. The full space of the cylindrical functions now becomes just space spanned by almost periodic functions and can be identified with the space of square summable functions on the Bohr compactification of the real line with the Haar measure $d\mu$.
\begin{equation}
  \mathcal{H}_{\bf kin} = L^2(\bar{\mathbb{R}},d\mu) 
\end{equation}

As a basis, eigenstates of the unit flux operator $\hat{p}$ can be selected. Such state can be labelled by a real number $\mu$ and the scalar product between its basis elements is discrete, namely:
\begin{equation}
  \braket{\mu_1|\mu_2}=\delta_{\mu_1,\mu_2}
\end{equation}
Thanks to that quantum vacuum is distinguished as $\delta_{\mu,0}\mathbb{I}$. Note that despite the discrete inner product $\mu$ can take any real value, which means that our Hilbert space is non-separable.
State in that base is expressed as: $\ket {\Psi}= \sum_i c^{(i)}\ket{\mu_i}$ with inner product: $\braket{\Psi_1|\Psi_2}=\sum_i \bar{c}^{(i)}_1c^{(i)}_2$. 
As the basic operators alongside $\hat{p}$, the $U(1)$ component of the holonomy has been chosen:
\begin{align}
\hat{p}\ket {\mu} =\frac{8\pi\gamma l_p^2}{6}\mu \ket {\mu}
~~~~~~~~&~~~~~~~~
\hat{e^{\frac{i\lambda c}{2}}}\ket {\mu} = \ket {\mu+\lambda}
\end{align} 
However, it is more convenient to work with slightly differently relabelled base and specifically chosen holonomy components. Namely  $\mu$ is exchanged to $v$ and restrict $\lambda$ to $\bar{\mu}(p)$ such that:
\begin{align}
\hat{V}\ket v &=\alpha |v| \ket v
&
\hat{N}\ket v :=\widehat{e^{\frac{i\bar{\mu} c}{2}}}\ket{v} = \ket {v+1}
\label{intv}
\end{align}
where $\braket{v|v'}=\delta_{v,v'} $ and $\alpha=2\pi G \hbar \gamma \sqrt{\Delta}$. The meaning of this choice will become clear later. This choice is tailored to the Hamiltonian constraint regularization in the so-called \emph{improved dynamics} scheme \cite{Ashtekar:2006wn}.
 

In the next step of the Dirac program, the constraints should be implemented. Due to symmetries and partial gauge fixing the Gauss and Diffeonorphism one are already satisfied. The only remaining is Hamiltonian one:
 \begin{equation}
H_{g}=H^E-2(1+\gamma^2)T
\label{equation 3}
\end{equation}
with Euclidean Hamiltonian and Lorentzian Hamiltonian constraint define as:
\begin{align}
H^E &= 
  \frac{1}{2\kappa}\int_\mathcal{V}d^3x\epsilon_{ijk}\frac{E^{ai} E^{bj}}{\sqrt{\det(h)}}F^k_{ab} 
 &
T &= \frac{1}{2\kappa}\int_\mathcal{V}d^3x\frac{E^{ai} E^{bj}}{\sqrt{\det(h)}}K^j_{[a}K^i_{b]}
\label{a1w}
\end{align}
where $h$ denotes metric on one 3-dimensional space-like surface $\Sigma_t$, $F^k_{ab}$ is curvature of connection $A^k_a$ and $K^i_a$ is extrinsic curvature 1-form. To be allowed to quantize constraint it must be re-expressed in terms of holonomies and densitized triads.
As suggested in \cite{Ashtekar:2006uz,Ashtekar:2006wn}, the term involving triads is re-expressed as:
\begin{equation}
\epsilon_{ijk}\frac{E^{ai} E^{bj}}{\sqrt{det(h)}}=\sum_k \frac{4sgn(p)}{\kappa\lambda V_0^{\frac{1}{3}}}\epsilon^{abc}\delta^k_c Tr\Big( h_i^{(\lambda)}\{(h_i^{(\lambda)})^{-1},V\}\tau_i \Big)
\label{equ12}
\end{equation}
and if we consider square loop $\square_{ij}$ (parallel to one of the faces of the elementary cell and with sides of length $\lambda V_0^{\frac{1}{3}} $ each)  the curvature of connection becomes: 
\begin{equation}
F^k_{ab}=-2 \lim_{Ar_{\square}\rightarrow 0}
Tr
\Big(
\frac{h^{(\lambda)}_{\square_{ij}}-1}{\lambda^2 V_0^{\frac{2}{3}}}
\Big)
\tau^k\delta^i_a\delta^j_b
\label{equA}
\end{equation}
where $h^{(\lambda)}_{\square_{ij}}$ is holonomy around square loop $\square_{ij}$, defined as:
\begin{equation}
h^{(\lambda)}_{\square_{ij}}=h_i^{(\lambda)}h_j^{(\lambda)}(h_i^{(\lambda)})^{-1}(h_j^{(\lambda)})^{-1}
\end{equation}

However classical limit $Ar_{\square}\rightarrow 0 $ is problematic in quantum theory - it means that loops area shrinks to be zero which, due to discontinuity, is forbidden in LQG. Rather than this, we set the loop till its area will be equal to twice the smallest non-zero eigenvalue of the area operator from LQG. In LQC it is called area gap $\Delta = 2A_1$:
\begin{equation}
\Delta=\bar{\mu}^2|p|\label{delta}
\end{equation}
One can now see that the new label $v$ introduced in \eqref{intv} corresponds to 'would be' affine parameter generated by the holonomy component $e^{i\frac{\bar{\mu}}{2}c}$ such that its quantum counterpart becomes a unit shift.

So by taking this into account and by plugging formula \eqref{equ12} and \eqref{equA} to Euclidean part of Hamiltonian constraint, it becomes in the desired form:
\begin{equation}
H^E=-\frac{2sgn(p)}{\kappa \gamma \bar{\mu}^3} \sum_{ijk} \epsilon^{ijk}Tr(h_i^{(\bar{\mu})}h_j^{(\bar{\mu})}(h_i^{(\bar{\mu})})^{-1}(h_j^{(\bar{\mu})})^{-1}h_k^{(\bar{\mu})}\{(h_k^{(\bar{\mu})})^{-1},V\})
\label{euclidean}
\end{equation}
Now we wish to reformulate the Lorentzian part of the constraint. And we will do it in a particular way by pointing out that on classical level $K^i_a$ can be rewritten, according to equation \eqref{taga5}, as $K^i_a=\frac{1}{\gamma}(A^k_a-\Gamma^i_a)$ to obtain:
\begin{equation}
E^{ai} E^{bj}K^j_{[a}K^i_{b]}=\frac{1}{2\gamma^2}\epsilon_{ijk}E^{ai} E^{bj}(F^k_{ab}-\Omega^k_{ab})
\label{omega}
\end{equation}
with $\Omega^k_{ab}$ define as curvature of spin connection $\Gamma^i_a$. Thanks to that, that Lorentzian part is constituted by two terms: one proportional to the Euclidean part and one corresponds to Ricci curvature. But for flat model $\Omega^k_{ab}=0$ thus Lotentzian part is only proportional to the Euclidean part and can be subsumed into it. Such procuder leads to the standard LQC presented in \cite{Ashtekar:2006uz, Ashtekar:2006wn}. However such a choice is not unique - different choices of expressing $K^i_a$ cause different operators in quantum theory - we will use that fact in the next sections.

%
 
%
%
Thus Hamiltonian constraint for standard LQC, takes the form:
\begin{equation}
H_g=-\frac{4sgn(p)}{\kappa \gamma \bar{\mu}^3} \sum_{ijk} \epsilon^{ijk}Tr(h_i^{(\bar{\mu})}h_j^{(\bar{\mu})}(h_i^{(\bar{\mu})})^{-1}(h_j^{(\bar{\mu})})^{-1}h_k^{(\bar{\mu})}\{(h_k^{(\bar{\mu})})^{-1},V\})
\label{equHGclasical}
\end{equation}
which after quantization (for improved dynamics scheme \cite{Ashtekar:2006wn}) becomes:
\begin{equation}
\hat{H_g}=\frac{24i}{\kappa^2 \gamma \hbar \sqrt{\Delta^3}} \hat{\sin}(b)\hat{A}\hat{\sin}(b)
\label{equHG}
\end{equation}
where:
\begin{align*}
\hat{A}=\hat{sgn}(v)\hat{V}\Big(\hat{\sin}(\frac{b}{2})\hat{V}\hat{\cos}(\frac{b}{2})-\hat{\cos}(\frac{b}{2})\hat{V}\hat{\sin}(\frac{b}{2})\Big) ~~,&~~\hat{\sin}(\frac{b}{2})=\frac{1}{2i}(\hat{N}-\hat{N}^{-1}) ~~,&~~ \hat{\cos}(\frac{b}{2})=\frac{1}{2}(\hat{N}+\hat{N}^{-1})
\end{align*}
and $b:=c\bar{\mu}$. This operator is selfadjoint and positive define on $\mathcal{H}_{\bf kin} = L^2(\bar{\mathbb{R}},d\mu) $. 
Its action on the state in volume basis returns:
\begin{equation}
\hat{H_g}\Psi(v)=f_{+}\Psi(v+4)+f_0\Psi(v)+f_{-}\Psi(v-4)
\end{equation}
 As a regular difference operator it divides its domain onto subspaces of functions supported on the lattices:
\begin{equation}
\mathcal{L}_{\epsilon}=\{ \epsilon+4n;~~ n \in \mathds{Z} \}
\end{equation}
where the value of $\epsilon\OG{\in [0,4]}$ can be fixed. Consideration of $\hat{H}_g$ can then be restricted only to Hilbert space supported on a chosen subset (lattice). Such subspaces, unlike the full Hilbert space, are separable. In our calculations, we will focus on a particular lattice $\epsilon=0$. In such case the subsets of $\mathcal{L}_{\epsilon}$ corresponding to $n \in \{0\}, \mathbb{Z}^{\pm}$ are not mixed by the action of $\hat{H}_g$ and we can work with just one semi-lattice.

Lets now focus on the matter content. We consider  free massless scalar field $\phi$ that is minimally coupled to gravity. So total Hilbert space given by:
\begin{equation}
\mathcal{H}= \mathcal{H}_{\bf kin} = L^2(\bar{\mathbb{R}},d\mu) \otimes L^2(\mathds{R}, d\phi)
\end{equation}
is build of solutions to total Hamiltonian constraint define as a tensor product of presented earlier (gravitational) Hamiltonian constraint \eqref{equHG} and constraint of the matter field which is provided by:
\begin{equation}
C^{\prime}_{\phi}= |p|^{-\frac{2}{3}}\frac{p^2_{\phi}}{2}
\end{equation}
Its basic Poisson bracket is given by
$
\{\phi, p_{\phi}\}=1
$
and Hamiltonian's equations are:
\begin{align}
\label{phiandphi}
&\dot{p_{\phi}} = \{ p_{\phi} , C^{\prime}_{\phi} \} = 0 ~~~~&~~~~
&\dot{\phi} = \{ \phi , C^{\prime}_{\phi} \} = |p|^{-\frac{2}{3}}p_{\phi} 
\end{align}
What can be read from them is that, $p_{\phi}$ is a constant of motion, thus $\phi$ is a monotonic function of the time and it will serve as a parameter with respect to which the whole system will be evolving. The procedure to quantize this constraint will be, to begin with,  $|p|^{-\frac{2}{3}}$ then (as in the standard quantum mechanic) $p_{\phi}$. But total Hamiltonian constraint can be multiplied by lapse function $N=2V$ (because it is a Lagrange multiplier it will not change physics). Thus in $C_{\phi}:=2VC^{\prime}_{\phi}$ only momentum of $\phi$ is quantized and constraint will be promoted to the following operator:
\begin{equation}
\hat{C}_{\phi}=\mathds{1}\otimes(-\hbar^2\partial_{\phi}^2)
\end{equation}
But rather than simply multiply the gravity part of the constraint with $N$, symmetric ordering with respect to the volume operator should be applied. It leads to the operator:
\begin{equation}
\hat{\Theta}=\frac{48i}{\kappa^2 \gamma \hbar \sqrt{\Delta^3}}\sqrt{\hat{V}} \hat{\sin}(b)\hat{A}\hat{\sin}(b)\sqrt{\hat{V}}\otimes\mathds{1}
\end{equation}
 
Thanks to all of that, full quantum constraint takes the form of Klein-Gordon equation:
\begin{equation}
\hbar^2\partial_{\phi}^2\Psi(v,\phi)=-\hat{\Theta}\Psi(v,\phi)
\label{phi}
\end{equation}
and allows to apply the standard quantum mechanical methods for finding solution (to a particular lattice $\epsilon=4$):
\begin{equation}
\Psi(v,\phi)=\int_{0}^{\infty}dk\Big(\Psi_+(k)e_k(v)e^{i\omega\phi}+\Psi_{-}(k)\bar{e}_k (v)e^{-i\omega\phi} \Big) 
\end{equation}
where $e_k(v)$ is eigenfunction for the operator $\Theta$ which corresponds to eigenvalue $\omega$ of operator of the matter field and $k$ is connected to it by $\omega^2=12\pi G k^2$. Such structure leads to another splitting on $\mathcal{H}_{\epsilon}$  corresponding to positive and negative energies. If we restrict consideration, for example for a sector of positive energies evolution of the system is provided by:
\begin{equation}
\Psi(v,\phi)=e^{i(\phi-\phi_0)\sqrt{\Theta}}\Psi(v,\phi_0)
\end{equation}
and the physical inner product for any given $\phi_0$ is:
\begin{equation}
\braket{\Psi_1|\Psi_2}_{\epsilon}=\sum_{v\in \mathcal{L}_{\epsilon}}\Psi_1(v,\phi_0) \bar{\Psi}_2(v,\phi_0)
\label{scaalrproductv}
\end{equation}

To complete the Dirac program, one more step is required to be done - the construction of physical observables. In the presented model they are found to be an operator of the momentum of the field $\phi$ and volume operator for fixed $\phi_0$ \cite{Ashtekar:2006uz, Ashtekar:2006wn}, which action on the state $\Psi$ returns:
\begin{align}
\hat{V}_{\phi_0}\Psi(v,\phi)&=\alpha e^{i(\phi-\phi_0)\sqrt{\Theta}}|v|\Psi(v,\phi_0) & \hat{p}_{\phi}\Psi(v,\phi)=-i\hbar\partial_{\phi}\Psi(v,\phi)
\label{diraac}
\end{align}
and they expectation values are:
\begin{align}
\braket{\hat{V}_{\phi_0}}&=\alpha \sum_{v\in \mathcal{L}_{\epsilon}}|v||\Psi(v,\phi_0)|^2 &\braket{\hat{p}_{\phi}}=-i\hbar \sum_{v\in \mathcal{L}_{\epsilon}}\bar{\Psi}(v,\phi) \partial_{\phi}\Psi(v,\phi)
\end{align}
To interpret the evolution of the model let's recall results from \cite{Ashtekar:2006wn} for the expectation value of the volume operator. Because General Relativity is recovered from LQC for a $\phi\rightarrow\pm\infty$ expectation values of volume  (by taking into account dispersion)  approach classical trajectories asymptotically. Departure becomes significant near-Planckian energy densities - classical trajectories collapsing into a singularity, while effects of quantum geometry cause repulsive behaviour of gravitational force leading to quantum bounce at critical energy density $\rho_c=\frac{3}{\kappa\gamma^2\Delta}$.

Similar results are obtained by examination of the effective dynamic of the system. Consideration begins with a regularized  but classical version of the full constraint \cite{Ashtekar:2006uz, Li:2018fco} given by: 
\begin{equation*}
H_{eff}=-\frac{3v\sin^2(b)}{\kappa\gamma^2\Delta}+\frac{p^2_{\phi}}{2v}
\end{equation*}
and because $H_{eff}=0$, the energy density $\rho=\frac{p^2_{\phi}}{2v^2}$ is: 
\begin{equation}
\rho=\rho_{c}\sin^2(b)
\label{rho}
\end{equation}
Evolution of this system is provided with Hamilton’s equation for $b$ and $v$ which takes form:
\begin{subequations}
\begin{align}
\label{vdot}
\dot{v}=&\{H_{eff}, v\}=\frac{3v}{\gamma\Delta}\sin(2b)\\
\dot{b}=&\{H_{eff}, b\}=-\frac{3v}{\gamma\Delta}\sin^2(b)+4\pi G\rho
\end{align}
\end{subequations}
and $\phi$ and it conjugate momentum as in \eqref{phiandphi}. By combining \eqref{vdot} with \eqref{phiandphi} and by using relation \eqref{rho} evolution of volume with respect to scalar field is expressed as:
\begin{equation}
\frac{dv}{d\phi}=\pm\sqrt{6\kappa(1-\frac{\rho}{\rho_c}})v
\end{equation}
which is satisfied, according to \cite{Ashtekar:2007em}, by hyperbolic function expressed as:
\begin{equation}
V= \frac{p_{\phi}}{\sqrt{2\rho_c}}\cosh(\beta\phi-\beta\phi_0)
\label{vmijn}
\end{equation}
This solution is in good agreement with quantum predictions.

\subsection{Strict Thiemann regularization of Loop Quantum Cosmology}\label{mlqc}

In Loop Quantum Cosmology we chose to at first, reformulate the
Lorentzian part to form, where it can be subsumed into the Euclidean
part and then to perform quantization of the Hamiltonian Constraint
given by formula \eqref{equHGclasical}. But as has been already pointed we are allowed to use different regularization formula to obtain constraint in elementary variables. For example in \cite{Yang:2009fp} two-approach were proposed, to which by following \cite{Saini:2019tem} we will refer as mLQC-I and mLQC-II. The first follows directly the algorithm proposed
originally by Thiemann \cite{Thiemann:2007zz} for full LQG namely, we obtain $K^i_a$ by a Poisson bracket:
\begin{equation}
K^i_a=\frac{1}{\kappa \gamma^3}\{A^i_a, \{H^E,V\}\}
\label{kia}
\end{equation}
which in cosmological setings reduce to:
\begin{equation}
K^i_a=-\frac{1}{\kappa
\gamma^3\lambda}h_i^{(\lambda)}\{(h_i^{(\lambda)})^{-1}, \{H^E,V\}\}
\end{equation}
To formulate above expresion in improved dynamics scheme \cite{
Ashtekar:2006wn} trivial swap from $\lambda$ to $\bar{\mu}$ is not
sufficient because of dependence of $\bar{\mu}$ on triads
\cite{Yang:2009fp} - formula \eqref{delta}. This leads to the following:
\begin{equation}
K^i_a=-\frac{2}{3\kappa
\gamma^3\bar{\mu}}h_i^{(\bar{\mu})}\{(h_i^{(\bar{\mu})})^{-1}, \{H^E,V\}\}
\end{equation}
and by plugging it - alongside with formula \eqref{equ12} - into
Lorentzian constraint, $T$ can be
expressed in required form \cite{Yang:2009fp}:
\begin{equation}
T = \frac{8sgn(p)}{\kappa^4 \gamma^7 \bar{\mu}^3} \sum_{ijk}
\epsilon^{ijk} \Tr\Big(h_i^{(\bar{\mu})}\{(h_i^{(\bar{\mu})})^{-1},
\{H^E,V\}\}h_j^{(\bar{\mu})}\{(h_j^{(\bar{\mu})})^{-1}, \{H^E,V\}\}
h_k^{(\bar{\mu})}\{(h_k^{(\bar{\mu})})^{-1},V\}\Big)
\end{equation}
The implementation of the Dirac program is the same as in the previous section. To be exact  - quantization on the kinematical level as well as the introduction of the matter field is the same as for mainstream LQC. The point of divergence is a slightly different form of the quantum Hamiltonian
constriant which Lorentzian part is:

\begin{equation}
\hat{T} = -\frac{96i}{9\kappa^4
\gamma^7\hbar^5}\Big(\hat{\sin}(\frac{b}{2})\hat{B}\hat{\cos}(\frac{b}{2})-\hat{\cos}(\frac{b}{2})\hat{B}\hat{\sin}(\frac{b}{2})\Big)\hat{A}\Big(\hat{\sin}(\frac{b}{2})\hat{B}\hat{\cos}(\frac{b}{2})-\hat{\cos}(\frac{b}{2})\hat{B}\hat{\sin}(\frac{b}{2})\Big)
\end{equation}
where: 
\begin{align}
\hat{\sin}(\frac{b}{2})=\frac{1}{2i}(\hat{N}-\hat{N}^{-1}) ~~&~~ \hat{\cos}(\frac{b}{2})=\frac{1}{2}(\hat{N}+\hat{N}^{-1})
\label{sincos}
\end{align}
and $\hat{B}$ is commutator between $\hat{H}^E$ and $\hat{V}$, while
Euclidean term is as in equation \eqref{euclidean}. Full Hamiltonian constraint again takes the form of the Klein-Gordon equation as in \ref{phi} but with new $\Theta$ formulated in shift and volume operators as: 
\begin{equation}
\hat{\Theta}=\frac{3}{2\kappa\hbar^2\Delta}\sqrt{\hat{V}}\Big(
-s'\hat{N}^4\hat{V}\hat{N}^4+\hat{N}^2\hat{V}\hat{N}^2+2(s'-1)\hat{V}+\hat{N}^{-2}\hat{V}\hat{N}^{-2}-s'\hat{N}^{-4}\hat{V}\hat{N}^{-4}
  \Big)\sqrt{\hat{V}}\otimes \mathds{1}
\end{equation}
where $s'=\frac{1+\gamma^2}{4\gamma^2}$. Thanks to that, it is
straightforward to see that, as in mainstream LQC, $\Theta$ acts on
states in volume representation as a difference operator - but now it is of the 4th order. So we again make a restriction to the superselection
sector for lattice $\epsilon=4$. But by defining Fourier transformation of the state in volume base to
momentum $b$ base as:
\begin{equation}
\psi(b)=\sum_{v\in\mathcal{L}_{\epsilon}}\frac{1}{\sqrt{|v|}}\psi(v)e^{i\frac{vb}{2}}
\label{fb1}
\end{equation}
with $b\in\left[ 0, \pi\right]$, $\hat{\Theta}$ becomes differential
operator of the second-order \cite{Assanioussi:2019iye}, given by the
formula:
\begin{equation}
\hat{\Theta}=12 \pi G \hbar^2 \gamma ^2 \Big((
\sin(b)\partial_b)^2-s^{\prime}(\sin(\frac{b}{2})\partial_b)^2\Big)
\end{equation}
which is in agreement with the order of the difference operator in
mainstream LQC. In addition,
$\hat{\Theta}$ is symmetric but not essentially self-adjoint. By
investigating its deficiency subspace which has both dimensions equal to
one (it was shown in \cite{Assanioussi:2019iye}), this operator admits a
family of self-adjoint extensions. It indicates that evolution is not unique and we can restrict ourselves to work on a subspace of $\mathcal{H}$ corresponding to a particular extension of $\hat{\Theta}$. To do so in addition some boundary data (corresponding to our choice) needs to be added. Then physical states are constructed by analogy to mainstream LQC but with basis build out of eigenfunctions evaluated for particular extension (it was done analytically in b-representation in \cite{Assanioussi:2019iye}).
 Moreover Klein Gordon equation
for mLQC-II for large volumes of the
The universe is stable (in the sense of Von Neumann stability
\cite{Saini:2019tem}). It means that eigenfunction
can be numerically evaluated from initial data in $v$.

\indent To check the dynamics physical observables
must be built. In LQC they are the momentum of the field $\phi$  and
volume at any given value of $\phi$. In this model, momentum can be
again promoted to the operator, but $\hat{V}_{\phi_0}$ could act outside of the
physical Hilbert space \cite{Assanioussi:2019iye, Pawlowski:2011zf}. Solution her is to simply
replace $\hat{V}_{\phi_0}$ with a bounded function of this operator.
Thus we recall from \cite{Assanioussi:2019iye}, as a second Dirac
observable - compactified volume:
\begin{equation}
\hat{\theta}_K=\arctan\Big(\frac{\hat{V}}{\alpha K}\Big)
\end{equation}
where $K$ is any positive real number. This operator is bounded by $\frac{\pi}{2}$ which corresponds to infinite volume obtained in finite value of scalar field - no such thing happens in considered earlier model. But to extract physics and not
exactly qualitative solution from it we can examine effective dynamic:
\begin{equation}
H_{eff}=\frac{3v}{\kappa\Delta}\Big(\sin^2(b)-s' \sin^2(2b)\Big)
\end{equation}
Now vanishing of constraint leds to two solutions:
\begin{equation}
\sin^2(b_{\pm})=\frac{1\pm\sqrt{1-\frac{\rho}{\rho_c^I}}}{2(\gamma^2+1)}
\end{equation}
where critical energy density in mLQC-I is:
\begin{equation}
\rho_c^I=\frac{\rho_c}{4(\gamma^2+1)}
\end{equation}
Solutions to $b_+$ and $b_-$ are both physically achievable and indicate asymmetric evolution of the universe with respect to the bounce. But more detailed analysis \cite{Assanioussi:2019iye}, shows that in fact for $ \phi \rightarrow \pm \infty $ General Relativity is recovered and the expectation value of volume, classical and effective trajectories will coincide. But in the near Planck energy regime classical trajectory will collapse into singularity while in quantum theory and effective dynamics bounce occurs. But after the bounce in the mLQC-I Universe will rapidly expand (as in GR in presence of cosmological constant) to a moment when the finite value of scalar field volume will be infinity. This point was identified as a transition from expanding Universe (in its conformal future) to contracting Universe (explicitly to its conformal past) where second bounce occurs \cite{Assanioussi:2019iye}. Thanks to that we can identify solutions to $b_+$
and $b_-$  (from effective dynamics) as corresponding to emerging DeSiter epochs (between bounces) and for trajectories for $ \phi \rightarrow \pm \infty $ respectively.

\section{Second Modification of Loop Quantum Cosmology}

In this chapter, we describe the next version of the LQC as a result of
Thieman's regularization, which (unlike the previous two)
has not been yet investigated in detail. It was first proposed (together with mLQC-I) in \cite{Yang:2009fp} and its effective dynamic has ben probed for example in \cite{Li:2018fco}. Because Dirac program of quantization is as in mainstream LQC and mLQC-I, in the section \ref{thetaV} we start at the point of divergence - construction of the constraint. To be exact we will present
regularization and action of operator $\Theta$ on the state in volume
representation and study those properties that can be probed
in this representation. Since for mLQC-I analysis in the momentum
representation has been much more useful, in the next section -
\ref{thetaB} we will perform a transformation to this representation from volume representation
and check properties of $\Theta$. In the
section \ref{dynamics} we will attempt to probe the quantum
dynamics via finding the expectation value of the volume as a
function of the scalar field and constant of motion. Finally in sec. \ref{effdynamics} we
will compare our results against obtained with effective dynamics.

\subsection{Regularization and $\hat{\Theta}$ in volume representation}\label{thetaV}

\begin{figure}[h]
\center
\includegraphics[scale=0.85]{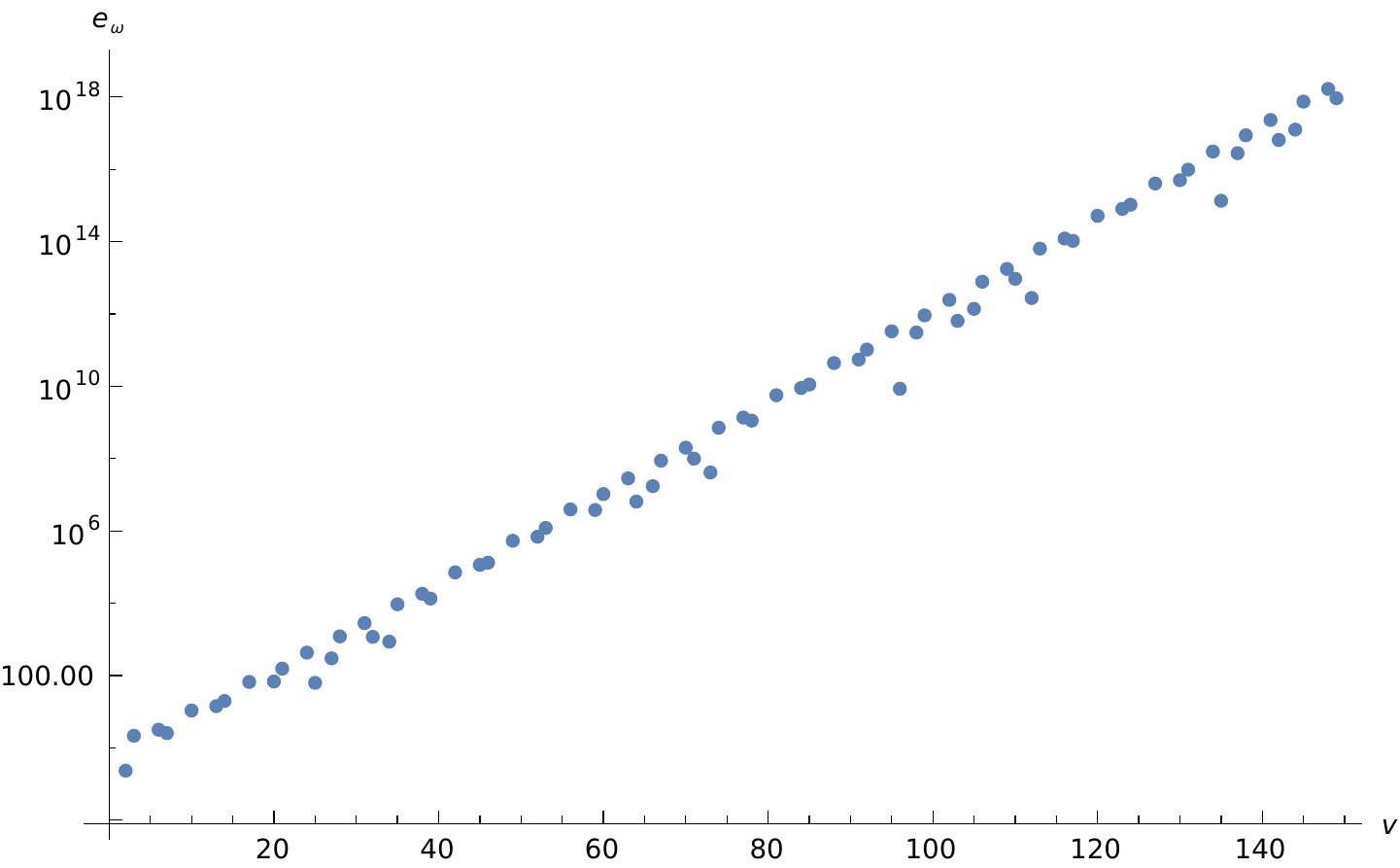}
\caption{Values of the eigenfunction $e_{\omega}$ to $\Theta$ -in
logarytmic scale, for  $\omega=10$}
\label{figa3}
\end{figure}

As mLQC-I quantization arises from taking the general form of $K^i_a$ in
The Lorentzian term, the second modification of LQC will be applied only to
spatially flat models because in that case extrinsic curvature 1-form,
can be expressed as:
\begin{equation}
K^i_a=\frac{1}{\gamma}A^i_a
\label{KIA}
\end{equation}
and constraint becomes:
\begin{equation}
T=\frac{1}{2\kappa\gamma^2}\int_\mathcal{V}d^3x\frac{E^{ai}
E^{bj}}{\sqrt{det(h)}}A^j_{[a}A^i_{b]}
\end{equation}
where at clasical level it is equivalent to $T$ formulated in equation
\eqref{a1w}. In cosmological setings, it reduce to:
\begin{equation}
T=-\frac{2sgn(p)}{\kappa^2\gamma^3}\epsilon^{ijk}\Tr\Big(c\tau_ic\tau_j\{c\tau_k,V\}\Big)
\end{equation}
where to obtain it in terms of holonomy we will use the following two
properties from \cite{Yang:2009fp}:
\begin{subequations}
\begin{align}
&c\tau_i=\lim_{\bar{\mu}\rightarrow
0}\frac{1}{2\bar{\mu}}\Big(h_i^{(\bar{\mu})}-(h_i^{(\bar{\mu})})^{-1}\Big)\\
&\{c\tau_k,V\}=h_k^{(\bar{\mu})}\{(h_k^{(\bar{\mu})})^{-1},V\}
\end{align}
\end{subequations}
and thus Lorentzian part becomes:
\begin{equation}
T = \frac{sgn(p)}{2\kappa^2 \gamma^3
\bar{\mu}^3}\lim_{\bar{\mu}\rightarrow 0} \sum_{ijk} \epsilon^{ijk}
\Tr\Big(h_i^{(\bar{\mu})}-(h_i^{(\bar{\mu})})^{-1})(h_j^{(\bar{\mu})}-(h_j^{(\bar{\mu})})^{-1})h_k^{(\bar{\mu})}\{(h_k^{(\bar{\mu})})^{-1},V\}\Big)
\end{equation}
And again as in mLQC-I quantization on the kinematical level is the same as in LQC thus we encounter the very same
problem in the above expression - ill-definiteness of the limit. To resolve this we proceed as already been mentioned - by setting the holonomy length the same way as when quantizing the Euclidean part of the Hamiltonian constraint. By this, the operator of the Lorentzian part becomes:
\begin{equation}
\hat{T}=-\frac{24i}{\gamma^3\kappa^2\hbar}\hat{\sin}(\frac{b}{2})\hat{A}\hat{\sin}(\frac{b}{2})
\end{equation}
where $\hat{\sin}(\frac{b}{2})$ is defined as in \eqref{sincos} and full operator of gravity part of Hamiltonian constraint in volume
representation can be defined as:
\begin{equation}
\hat{\Theta}=-3 \pi G \hbar^2 \gamma ^2
\sqrt{|\hat{v}|}\Big(\hat{N}^2|\hat{v}|\hat{N}^2-s\hat{N}|\hat{v}|\hat{N}+2(s-1)|\hat{v}|-s\hat{N}^{-1}|\hat{v}|\hat{N}^{-1}+\hat{N}^{-2}|\hat{v}|\hat{N}^{-2}\Big)\sqrt{|\hat{v}|}
\label{thetamlqc2v}
\end{equation}
where $s=\frac{4(1+\gamma^2)}{\gamma^2}$. In volume representation
$\hat{\Theta}$ act on the state as in LQC as difference operator and as
in mLQC-I it is of 4th order operator. If we consider its
eigenvalue problem (corresponding to an eigenvalue of matter part operator $\omega$), this
an equation can be rewritten as recurrence relation:
\begin{equation}
f_{-4}e_{\omega}(v+4)=f_{-2}e_{\omega}(v+2)+(\omega^2-f_0)e_{\omega}(v)-f_{+4}e_{\omega}(v-4)+f_{+2}e_{\omega}(v-2)
\end{equation}
where:
\begin{subequations}
\begin{align}
f_{+4} &= \sqrt{v}\sqrt{v-4}|v-2|
&
f_{-4} &= \sqrt{v}\sqrt{v+4}|v+2|
\\
f_{+2} &= s\sqrt{v}\sqrt{v-2}|v-1|
&
f_{-2} &=s\sqrt{v}\sqrt{v+2}|v+1|
\\
f_{0} &= 2(s-1)v^2
\end{align}
\end{subequations}
and solutions are obtained by fixing for $e_{\omega}(v=2)$ and
$e_{\omega}(v=4)$. This indicates that the space of solutions has
dimension 2. We performed a numerical analysis of this problem. What we have found is that solution is unstable - they will grow exponentially as it is shown (for particular eigenvalue $\omega=10$) in Fig.\ref{figa3}. It means that unlike in mLQC-I, eigenfunction in volume representation cannot be numerically evaluated from initial data. The instability of these solutions was first established via VonNeuman stability analysis in \cite{Saini:2019tem} where the large volume behavior of the equation has been studied. Here we presented the roots of this equation:
\begin{subequations}
\begin{align}
1,~~&~~ 1\\
\frac{2+\gamma^2-2\sqrt{1+\gamma^2}}{\gamma^2},~~&~~ \frac{2+\gamma^2+2\sqrt{1+\gamma^2}}{\gamma^2}
\end{align}
\end{subequations}
where first is made of two identical roots equal to one but in the second pair at least one is clearly greater than unity. 
It not necessary means that in large volumes of the Universe,
GR will not be recovered. Rather it points out the possibility
of the occurrence of not valid physical roots of the
equation. It was confirmed by investigation of the effective dynamics in \cite{Li:2018fco}, where one of the solutions to the vanishing of effective constraint was identified as nonphysical. Now to check properties of $\Theta$ we switch as in \cite{Assanioussi:2019iye} to momentum $b$ representation.

\subsection{$\hat{\Theta}$ in momentum $b$ representation}\label{thetaB}

\indent We begin with defining a type of a Fourier transformation
of state in volume base to momentum $b$ base as:
\begin{equation}
\psi(b)=\sum_{v\in\mathcal{L}_{\epsilon}}\frac{1}{\sqrt{|v|}}\psi(v)e^{i\frac{vb}{2}}
\end{equation}
wherewith respect to formula \eqref{fb1}, the novelty lies in the domain
of $b$ - now is a circle of unit radius rather than half unit.
We also write down, formula for the inverse of the above transformation:
\begin{equation}
\psi(v)=\frac{1}{\pi}\sqrt{|v|}\int_0^{2\pi}db\psi(b)e^{-i\frac{vb}{2}}
\end{equation}
and by plugging it in \eqref{scaalrproductv} scalar product in this
representation is:
\begin{equation}
  \braket{\psi_1|\psi_2}=\frac{1}{\pi^2}\sum_{v\in\mathcal{L}_{\epsilon}}|v|\int dbdb^{\prime} \bar{\psi}_1(b^{\prime})e^{i\frac{vb^{\prime}}{2}}\psi_2(b)e^{-i\frac{vb}{2}}
\end{equation}

where because of the absolute value of volume, $\braket{\psi_1|\psi_2}$
cannot be formulated as a single integral on the whole Hilbert space.
But if we consider projections  $P^\pm$ such that $[P^\pm\psi](v) = \theta(\pm
v)\psi(v)$ products on subspaces of positive and negative valuse of $v$ can be presented as:
\begin{align}
\nonumber
\braket{\psi_1|\psi_2}_{\pm}&=\pm\frac{1}{\pi^2}\sum_{v\in\mathcal{L}_{\epsilon}}\int
dbdb^{\prime}
\bar{\psi}_1(b^{\prime})v\psi_2(b)e^{-i\frac{v(b-b^{\prime})}{2}}=\pm\frac{1}{\pi}\int
dbdb^{\prime} \bar{\psi}_1(b^{\prime})(-2i\partial_b)\psi_2(b)\delta(b-b')\\
&=\mp\frac{2i}{\pi}\int db \bar{\psi}_1(b)\partial_b\psi_2(b)
\label{scalarb}
\end{align}
However, because we work with states symmetric in $v$ (thus need only positive semilattice) we will restrict ourselves to work with non-negative and grater
than 2 values of $v$ so we will be using integral with the minus sign.
Thanks to that, it is straightforward to find action of volume operator as:
\begin{align}
\alpha|\hat{v}|\psi(b)=-2i\alpha\partial_b \psi(b)
\end{align}
We now, by inspecting:
\begin{align}
\nonumber\sqrt{|\hat{v}|}\hat{N}\frac{1}{\sqrt{|\hat{v}|}}\psi(b)&=\sum_{v\in\mathcal{L}_{\epsilon}}\frac{1}{\sqrt{|v|}}\sqrt{|\hat{v}|}\hat{N}\frac{1}{\sqrt{|\hat{v}|}}\psi(v)e^{i\frac{vb}{2}}=\sum_{v\in\mathcal{L}_{\epsilon}}\frac{1}{\sqrt{|v|}}\frac{\sqrt{|v|}}{\sqrt{|v-1|}}\psi(v-1)e^{i\frac{vb}{2}}\\
&=\sum_{v'\in\mathcal{L}_{\epsilon}}
\frac{1}{\sqrt{|v'|}}\psi(v')e^{i\frac{v'b}{2}}e^{i\frac{b}{2}}=e^{i\frac{b}{2}}\psi(b)
\end{align}
find action of the shift operator to be in the following form:
\begin{equation}
\hat{N}\psi(b)=\frac{1}{\sqrt{|\hat{v}|}}\exp(\frac{ib}{2})\sqrt{|\hat{v}|}\psi(b)
\end{equation}
and thus:
\begin{equation}
\hat{\sin}(b)\psi(b)=\frac{1}{\sqrt{|\hat{v}|}}\sin(b)\sqrt{|\hat{v}|}\psi(b)
\end{equation}
In such case, $\hat{\Theta}$ takes the form:
\begin{align}
\nonumber\hat{\Theta}&=12 \pi G \hbar^2 \gamma ^2 \Big((
\sin(b)\partial_b)^2-\frac{4(1+\gamma^2)}{\gamma^2}(\sin(\frac{b}{2})\partial_b)^2\Big)\\
&=-12 \pi G \hbar^2 \gamma ^2\Big(f(b)\partial_b^2+h(b)\partial_b\Big)
\label{thetab1}
\end{align}
which is the differential operator of the second order where:
\begin{align}
f(b)=4\sin^2(\frac{b}{2})\Big(\frac{1+\gamma^2}{\gamma^2}-\cos^2(\frac{b}{2})\Big)~~&~~h(b)=\sin(b)\Big(\frac{1+\gamma^2}{\gamma^2}-\cos(b)\Big)
\end{align}
Because $f(b)$ is always non-negative, unlike in
\cite{Assanioussi:2019iye},  the signature of the Klein-Gordon equation - with $\Theta$ defined as in \eqref{thetab1} - will
not be mixed and constraint will be hyperbolic. Now, we wish to
introduce a new variable as a function of $b$ so that the
equation for evolution takes the strict K-G form.  After application of chain rule,
alongside with condition
\begin{equation}
f(b)(\frac{dx}{db})^2=1
\end{equation}
which has been reformulated to 
\begin{equation}
\frac{dx}{db}=\frac{1}{\sqrt{f(b)}}
\end{equation}
$x(b)$ has been be integrated out analytically, to be:
\begin{equation}
x(b)=\tanh^{-1}\Big(\frac{\sqrt{2}\cos(\frac{b}{2})}{\sqrt{2+\gamma^2-\gamma^2\cos(b)}}\Big)
\end{equation}
with $x(b=\pi)=0$ and appropriate limits:
\begin{align*}
  \lim_{b\rightarrow 0} x(b)=-\infty ~~~~~ &~~~~~ \lim_{b\rightarrow
2\pi}x(b)=\infty
\end{align*}
so, the range of this function is the whole real line. Moreover this
function is well defined  (continuous and differentiable at every point)
on the whole domain. Operator in new coordinates becomes:
\begin{equation}
\hat{\Theta}=-12 \pi G \hbar^2\partial_x^2
\end{equation}
and sclar product in $x$ takes form:
\begin{equation}
\braket{\psi_1|\psi_2}=-\frac{2i}{\pi}\int^{\infty}_{-\infty} dx
\bar{\psi}_1(x)\partial_x\psi_2(x)
\label{scalarx}
\end{equation}
The operator $|\hat{v}|$ on the subdomain supported on $v>0$  has been found after expressing $\frac{dx}{db}$ as a
function of $x$ variable:
\begin{equation}
|\hat{v}|=-i\frac{1}{\sqrt{\gamma^2+1}}f(x)\partial_x
\end{equation}
where $f(x)=(\gamma^2\tanh^2(x)+1)\cosh(x)$.\\
\indent Now we will check whether $\hat{\Theta}$ is self-adjoint
operator. To do so lets inspect the deficiency subspaces
$\mathcal{K}_{\pm} $ defined \cite{Reed:1975uy} as the spaces of
(kinematically) normalizable solutions $\psi^{\pm}$ to the equation:
\begin{equation}
12 \pi G \hbar^2 \partial_x^2\psi^{\pm}=\pm i\psi^{\pm}
\label{psiiplusminus}
\end{equation}
where both solutions have been found to be as plain weave type:
\begin{align}
\psi^{+} &= \exp(\frac{(-1)^\frac{1}{4}x}{\sqrt{12 \pi G}
\hbar})+\exp(-\frac{(-1)^\frac{1}{4}x}{\sqrt{12 \pi G} \hbar })
&
\psi^{-} &= \exp(\frac{(-1)^\frac{3}{4}x}{\sqrt{12 \pi G} \hbar
})+\exp(-\frac{(-1)^\frac{3}{4}x}{\sqrt{12 \pi G} \hbar })
\label{sol}
\end{align}
We now, as it was defined in \eqref{scalarx} computed
$\braket{\psi^{+}|\psi^{+}}$ and $\braket{\psi^-|\psi^{-}}$ to find out,
that both of these scalar products are not finite. It means that both
solutions \eqref{sol} are not normalizable and by that only zero will
satisfy equation \ref{psiiplusminus}. So both deficiency subspaces are
spaned only by a zeroth vector so:
\begin{align*}
\dim(\mathcal{K}_+)=0 ~~~~&~~~~ \dim(\mathcal{K}_{-})=0
\end{align*}
and by using the definition from \cite{Reed:1975uy} which was quoted in
Appendix B, we conclude that $\hat{\Theta}$ is essentially self-adjoint.
 
\subsection{Dynamics}\label{dynamics}
Because $\Theta$ turns out to be essentially self-adjoint evolution of the system is uniquely determined. As well physical Hilbert space can be identified as a solution to full quantum constraint. Thus we start probing the dynamic by writing  it  in $x$ representation:

\begin{equation}
\partial_{\phi}^2\Psi(x,\phi)=12\pi G\partial_{x}^2\Psi(x,\phi)
\label{cons}
\end{equation}
Eigenfunctions of $\partial_{\phi}^2 $ and $\partial _{x}^2$ are
$e^{ikx}$, $e^{i\omega\phi}$ respectively where $\omega$ is
,,frequency'' and it is connected to $k$ by $\omega=\sqrt{12\pi G}|k|$.
Solutions to equation \eqref{cons}, are find by using
standard quantum mechanical methods to be:
\begin{equation}
\Psi(x,\phi)=\int_{0}^{\infty}dk\psi(k)e^{ikx}e^{i\beta|k|\phi}
\label{phixk}
\end{equation}
where $\beta=\sqrt{12\pi G}$. Finally by complete
analogous to formulas \eqref{diraac}, Dirac observables are:
\begin{align}
\hat{V}_{\phi_0}\Psi(x,\phi)=\alpha
e^{i(\phi-\phi_0)\sqrt{\Theta}}|v|\Psi(x,\phi_0) ~~~~~~&~~~~~~
\hat{p}_{\phi}\Psi(x,\phi)=-i\hbar\partial_{\phi}\Psi(x,\phi)
\label{Diracsmlqc2}
\end{align}
Now we wish to find the expectation value of $V$ in $x$ representation. So as
mentioned earlier, we restricted ourselves to act only on subspace where
$v$ is nonnegative and larger than $2$. It leads to:
\begin{equation}
\braket{\psi_1|\hat{V}\psi_2}=-\frac{2\alpha}{\pi}\frac{1}{\sqrt{\gamma^2+1}}\int^{\infty}_{-\infty}
dx \bar{\psi}_1(x)(f(x)\partial^2_x +f'(x)\partial_x)\psi_2(x)
\label{vba}
\end{equation}


Now by using form \eqref{phixk} of the wave function in the above
expression we have:
\begin{align}
\nonumber\braket{\psi_1|\hat{V}\OG{|_{\phi}}\psi_2}&=
-\frac{2\alpha}{\pi}\frac{1}{\sqrt{\gamma^2+1}}\int^{\infty}_{-\infty}\int^{\infty}_{0}\int^{\infty}_{0}
dxdkdk' \bar{\psi}_1(k')e^{-ik'(x+\beta\phi)}(f(x)\partial^2_x
+f'(x)\partial_x)\psi_2(k)\bar{\psi}_2(k)e^{ik(x+\beta\phi)} \\
\nonumber
&=-\frac{2\alpha}{\pi}\frac{1}{\sqrt{\gamma^2+1}}\int^{\infty}_{-\infty}\int^{\infty}_{0}\int^{\infty}_{0}
dxdkdk'\bar{\psi}_1(k')(-k^2f(x)
+ikf'(x))\psi_2(k)e^{i(k-k')x}e^{i(k-k')\beta\phi)}\\
&=-\frac{2\alpha}{\pi}\frac{1}{\sqrt{\gamma^2+1}}\int^{\infty}_{0}
dk\bar{\psi}_1(k)(-k^2f(x) +ikf'(x))\psi_2(k)
\label{f1x}
\end{align}
To proceed further we need function $f$
and $f'$ in energy $k$ variable. At first we expand them in to Taylor series:
\begin{equation}
f(x)=\sum_{n=0}^{\infty}\frac{1}{n!}f^n(0)x^n
\end{equation}
then by investigation of
\begin{align}
\nonumber \hat{x}\Psi(x,\phi)&=\int_{0}^{\infty}dk
x\psi(k)e^{ikx}e^{i\beta|k|\phi}\\
&=\int_{0}^{\infty}dk\psi(k)(i\partial_k-\beta\phi)e^{ikx}e^{i\beta|k|\phi}
\end{align}
we find that $x=i\partial_k-\beta\phi $ and thus by
using binomial formula:
\begin{equation}
x^n=(i\partial_k-\beta\phi)^n=\sum_{l=0}^{n}{n\choose
l}(-\beta\phi)^{n-l}(i\partial_k)^l
\end{equation}
so, finally, by detailed analysis function, $f$ can be
re-expressed as:
\begin{equation}
f(x)=\sum_{i=0}^{\infty}\frac{1}{l!}\partial^l_{\beta\phi}f(\beta\phi)(i\partial_k)^l
\end{equation}
By plugging this form of functions into formula \eqref{f1x} we ended with:
\begin{align}
\nonumber\braket{\hat{V}}&=\frac{2\alpha}{\pi}\frac{1}{\sqrt{\gamma^2+1}}\sum_{i=0}^{\infty}\frac{1}{l!}\partial^l_{\beta\phi}f(\beta\phi)\int^{\infty}_{0}
dk\bar{\psi}_1(k)k^2(i\partial_k)^l\psi_2(k)\\
&-\frac{2\alpha}{\pi}\frac{1}{\sqrt{\gamma^2+1}}\sum_{i=0}^{\infty}\frac{1}{l!}\partial^l_{\beta\phi}f'(\beta\phi)\int^{\infty}_{0}
dk\bar{\psi}_1(k)ik(i\partial_k)^l\psi_2(k)
\label{vink}
\end{align}
Before going further, we find form of the scalar product in $k$
representation, we start from formula \eqref{scalarb}:
\begin{align}
\nonumber\braket{\psi_1|\psi_2}&=-\frac{2i}{\pi}\int_{0}^{2\pi} db
\bar{\psi}_1(b)\partial_b\psi_2(b)\\
\nonumber&=-\frac{2i}{\pi}\int_{0}^{2\pi}\int_0^{\infty}\int_0^{\infty} db dk dk' \bar{\psi}_1(k')e^{-i\beta
k'\phi}e^{-ik'b}\partial_b \psi_2(k)e^{i\beta k\phi}e^{ikb}\\
\nonumber&=-\frac{2i}{\pi}\int_{0}^{2\pi}\int_0^{\infty}\int_0^{\infty} db dk dk' \bar{\psi}_1(k')ik
\psi_2(k)e^{i\beta (k-k')\phi}e^{ik(b-b')}\\
&=\frac{2}{\pi}\int_0^{\infty} dk \bar{\psi}_1(k)k \psi_2(k)
\end{align}
thus formula \eqref{vink} becomes:
\begin{equation}
\braket{\hat{V}}=\alpha\frac{1}{\sqrt{\gamma^2+1}}\sum_{l=0}^{\infty}\frac{1}{l!}\Big(\partial^l_{\beta\phi}f(\beta\phi)
\braket{:k(i\partial_k)^l:}-i\partial^l_{\beta\phi}f'(\beta\phi)\braket{(i\partial_k)^l}\Big)
\label{vhaa}
\end{equation}
 where $:\cdot:$ denotes some symmetric ordering.

Now let us consider a system that is completely described by the
position $X$ and its conjugate momentum $P$. During quantization, these
variables become operators and physics is extracted through their
expectation values. But because the expectation value of the product of
operators, in general, is not equal to products of expectation values,
to describe the system completely after quantization we need moments
\cite{Bojowald:2010qm, Bojowald:2005cw}, defined as:
\begin{align}
\nonumber G^{a,b}&:=\braket{(\hat{P}-P)^a(\hat{X}-X)^b}_{Weyl}\\
&=\sum_{k=0}^a\sum_{n=0}^b(-1)^{a+b-k-n}{a\choose k}{b\choose
n}P^{a-k}X^{b-n}\braket{\hat{P}^k\hat{X}^n}_{Weyl}
\label{gab}
\end{align}
where index on expectation value means totally symmetric ordering. After
inverting relation \ref{gab} to obtain the formula for
$\braket{\hat{P}^k\hat{X}^n}_{Weyl}$, we can apply it to find:
\begin{align}
\nonumber\braket{k(i\partial_k)^l}&=\sum_{n=0}^1\sum_{i=0}^l{l\choose
i}\braket{k}^{1-n}\braket{i\partial_k}^{l-i}G^{ni}\\
&=\braket{k}\braket{i\partial_k}^{l}+\braket{i\partial_k}^{l}G^{1,0}+\sum_{i=1}^l{l\choose
i}\braket{i\partial_k}^{l-i}\Big(\braket{k}G^{0,i}+G^{1,1}\Big)
\label{br}
\end{align}
where on the left-hand side of the equation \eqref{br}, $\braket{...}$
denotes expectation value while on the right-hand side, it represents
classical trajectories. Form of \eqref{br} indicates that leading term is
$\braket{k}\braket{i\partial_k}^l$ and terms with $G^{a,b}$ are quantum
corections. If they will be neglected:
\begin{equation}
\braket{k(i\partial_k)^l}\cong\braket{k}\braket{i\partial_k}^l
\end{equation}
and by plugging it to equation \eqref{vhaa}, we found semi classical
trajectory of volume to be:
\begin{equation}
\braket{\hat{V}}\cong\frac{\alpha}{\sqrt{\gamma^2+1}}\Big(\braket{k}f(\beta\phi+\braket{i\partial_k})-if'(\beta\phi+\braket{i\partial_k})\Big)
\end{equation}
But it is straightforward to see an alarming property of this solution -
because the second term is pure imaginary, the whole solution will
return a complex value. To resolve this problem one more comment on
\eqref{br} should be done - in this formula sum of all terms must be
real but each of them could be complex. It means if we take the whole
expression for $ \braket{k (i \partial_k)^l} $ imaginary part will
cancel out. So we can just delete the problematic part. Thanks to that:
\begin{equation}
\braket{\hat{V}}\cong\frac{\alpha}{\sqrt{\gamma^2+1}}\braket{k}\Big(\gamma^2\tanh^2(\beta\phi+\braket{i\partial_k})+1\Big)\cosh(\beta\phi+\braket{i\partial_k})
\label{vg}
\end{equation}
which is symmetric function.
\subsection{Effective Dynamics}\label{effdynamics}
Now we will check if our result is consistent with the prediction of
effective dynamics, which with good approximation provides quantitative
prediction about quantum system \cite{Ashtekar:2011ni}. To do so, we
will be treating full Hamiltonian constraint - after regularization - as
a classical constraint. In the following consideration we are inspecting
form of $H_{eff}$ from \cite{Li:2018fco} given by:
$$
H_{eff}=-\frac{3v}{\kappa \Delta
\gamma^2}\sin^2(\frac{b}{2})[1+\gamma^2\sin^2(\frac{b}{2})]+\frac{p^2_{\phi}}{2v}
$$
which condition that it must vanish leads to:
\begin{equation}
\rho=4\rho_c\sin^2(\frac{b}{2})\Big(1+\gamma^2\sin^2(\frac{b}{2})\Big)
\label{rhomlqc}
\end{equation}
with equation of motion \cite{Li:2018fco} given by:
\begin{subequations}
\begin{align}
\dot{b}&=\{b,H_{eff}\}=-\frac{6v}{\gamma\sqrt{\Delta}}\sin^2(\frac{b}{2})(1+\gamma^2\sin^2(\frac{b}{2}))+4\pi
G\rho\label{veff1}\\
\dot{v}&=\{v,H_{eff}\}=\frac{3v}{\gamma\sqrt{\Delta}}\sin(b)(1+2\gamma^2\sin^2(\frac{b}{2}))
\label{veff}
\end{align}
\end{subequations}
But derivative in above expressions are with respect to conformal time
and to compare the evolution of $v$ with one obtained in \eqref{vg} we need
to formulate it, as a function dependent on the scalar field. To do so we
introduce an auxiliary function define as:
\begin{equation}
y:=\sin^2(\frac{b}{2})
\end{equation}
and by using formula \eqref{veff1} we have found:
\begin{equation}
\frac{db}{d\phi}=\frac{db}{dt}(\frac{d\phi}{dt})^{-1}=-2\beta y
(1+\gamma^2y)
\end{equation}
Now we calculate:
\begin{equation}
(\frac{dy}{d\phi})^2=\frac{1}{4}y(1-y)(\frac{db}{d\phi})^2=\beta^2y^2(1-y)(1+\gamma^2y)
\end{equation}
what allow us to get $y$ as a function of scalar field:
\begin{equation}
y(\phi)=\frac{1-\tanh^2(\beta\phi-\beta\phi_0)}{\gamma^2\tanh^2(\beta\phi-\beta\phi_0)+1}
\end{equation}
and reformulated $\rho$ from \eqref{rhomlqc} as:
\begin{equation}
\rho=4\rho_c\frac{1+\gamma^2}{(\gamma^2\tanh^2(\beta\phi-\beta\phi_0)+1)^2\cosh^2(\beta\phi-\beta\phi_0)}
\end{equation}
and by thus we obtain:
\begin{align}
\nonumber V&=\frac{|p_{\phi}|}{\sqrt{2\rho}}=\frac{\hbar\sqrt{12\pi G}
|k|}{\sqrt{8\rho_c}\sqrt{(1+\gamma^2)}}(\gamma^2\tanh^2(\beta\phi-\beta\phi_0)+1)\cosh(\beta\phi-\beta\phi_0)\\
&=\frac{\alpha}{\sqrt{(1+\gamma^2)}}|k|(\gamma^2\tanh^2(\beta\phi-\beta\phi_0)+1)\cosh(\beta\phi-\beta\phi_0)
\end{align}
which, by defing $\phi_0=-\frac{1}{\beta}\braket{i\partial_k}$ is
exactly what we have acquired in formula \eqref{vg}. It means that our
solution is consistent with one obtained in \cite{Li:2018fco}. Morover
it indicates that:
\begin{itemize}
\item evolution of the Universe will be symetric with respect to the
bounce that will ocure for: $$\rho_c^{II}=4(1+\gamma^2)\rho_c$$
\item for $ \phi \rightarrow \pm \infty $ General Relativity will be recovered.
\end{itemize}

\section{Conclusion}
In this thesis, we recollected the most important information about two
regularization prescriptions studied in detail in literature - the so
called (mainstream) LQC and mLQC-I.
 As well, we provided our detailed analysis of prescription first introduced (alongside with mLQC-I) in \cite{Yang:2009fp} to which we referred as mLQC-II. Because quantization on the kinematical level is the same as in the prescription mentioned earlier, we have begun by building the gravity part of the constraint. To be exact we started with the Lorentzian part (because Euclidean is as in formula \eqref{euclidean}) and pointing out that because we chose to work in k=0 FLRW model externistic curvature 1-form can be expressed as a term proportional to Ashtekar-Barbero connection \eqref{KIA}.
Then we expressed this constraint in terms of elementary variables and performed quantization as in \cite{Yang:2009fp}. It has been done to obtain the full operator of gravity part as in \eqref{thetamlqc2v}. Because we chose to introduce the matter field as in mainstream LQC full quantum constraint forms a type of Klein-Gordon equation where the scalar field $\phi$ play the role of the evolution parameter. By investigation of properties of $\Theta$ in volume representation, we identified it as difference operator of the fourth-order and by reformulating its eigenvalue problem
to a recurrence relation, we find out that its solution cannot be evaluated numerically for given initial conditions in $v$. It means that solution is unstable which confirmed the conclusion from \cite{Saini:2019tem}. Because of that, we chose to examine the properties of the operator of the gravitational part of constraint in a more convenient representation. Thus, as suggested in \cite{Assanioussi:2019iye}, we switched to momentum $b$ representation in which $\Theta$ has been
found to act as a differential operator of the second order. At this
point, we inspected its deficiency subspace to determined its
self-adjointness.
What we have found is that both of these
subspaces are built only of zeroth vector which means that the operator is essentially self-adjoint. By this, evolution can be uniquely determined. Then we find physical Hilbert space constituted of states satisfying condition \eqref{cons} which has a form of standard Klein-Gordon equation. As well we have found Dirac observables given by formula \eqref{Diracsmlqc2}. Following that, we attempted to probe the dynamic of that system by evaluating the expectation values of the volume as a function of $\phi$ for semiclassical states. We found out that this function is symmetric and built out of hyperbolic functions. Moreover, our result is in agreement with results obtained in \cite{Li:2018fco}. It means that for $\phi\to\pm\infty$ classical behavior predicted by General Relativity will be recovered. Departures from classical theory, become at near-Planckian energy density regime, where for critical energy density $\rho_c^{II} \approx 1,73\rho_{Pl}$ bounce will occur.

\begin{figure}[h]
\center
\includegraphics[scale=0.89]{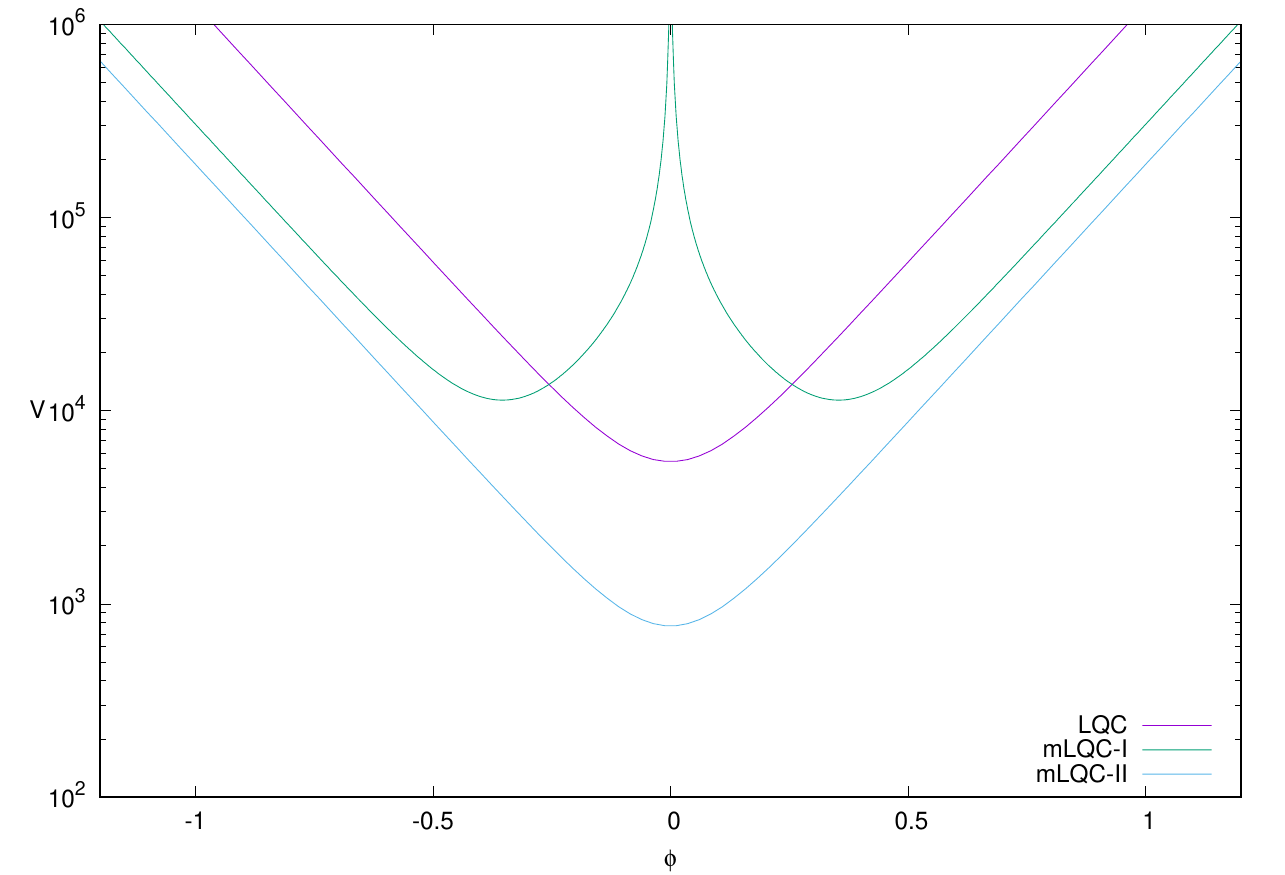}
\caption{Effective Dynamical Trajectories for $p_{\phi} = 5000$, Source: prof. Tomasz Pawłowski}
\label{figa4}
\end{figure}

We can now compare obtained results of mLQC-II against other
known regularizations. We can point out the following observations:
\begin{itemize}
\item In $\phi\to\pm\infty$ trajectories obtained through both effective dynamics 
and quantum methods for all three models approximate well provided than by General Relativity. It means that in the asymptotic future/past every known regularization of the Loop Quantum Cosmology will recreate classical theory. Effective trajectories has been gathered in Fig\ref{figa4} - for LQC it is described by hyperbolic cosine of $\phi$ \eqref{vmijn},  for mLQC-II it is again by hyperbolic cosine but multiplied by $(\gamma^2\tanh^2(\phi)+1)$ \eqref{vg} and for mLQC-II it has been obtained numerically.
\item In all models, the singularity of the Big Bang of General Relativity has been resolved.  At near-Planckian energy densities regime, we encounter bounce that occurs at a critical density which differs for each prescription and they satisfy:
$$
\rho_c^I < \rho_c < \rho_c^{II}
$$
\item As was shown for mLQC-I modifications are not necessarily restricted to
Planckian density regions - rather than one bounce two occurs. Between them emerging two DeSitter epochs separated by point were for a finite value of the scalar field, the volume of the Universe becomes infinite.
\item A different approach in the regularization of the gravity part of Hamiltonian constraint cause different forms and properties of operator $\Theta$. In mainstream LQC it is a self-adjoint (thus evolution can be uniquely determined) operator which acts on the state in volume representation as difference operator and in momentum $b$ representation as a differential operator. Both are of the second order. In mLQC-I differential, an operator is again of the second-order but the order of the difference one is four. That discrepancy is accounted for by
admitting multiple self-adjoint extensions, each generating independent
evolution. For mLQC-II we have a similar situation (fourth for $v$, second for
$b$), but both deficiency subspaces are
spaned only by a zeroth vector so the operator is essentially self-adjoint. This apparent paradox is resolved
by the fact that only some subspace of solutions to the eigenvalue problem
spans physical Hilbert space.
\end{itemize}


\subsection*{\centering{Acknowledgements}}
At first, I would like to express my gratitude to my supervisor prof. Tomasz Pawłowski without whom this work couldn't be done. Thank you for guiding me and helping to solve an encountered problems. I also want to thank all my lecturers and fellow students from Physics Department for encouraging me to continue study.
\newpage
\section{Plots}
~~\\
~~\\
\newpage

\begin{appendices}
\section{Constraints}
\label{appA}
Einstein equations (without cosmological constant) are obtained by variation (with respect to the metric tensor) of Einstein-Hilbert action:
\begin{equation}\tag{A.1}
S_{EH}=\frac{1}{2\kappa}\int_{\mathcal{M}}d^4xR\sqrt{-g}
\label{A.1.1}
\end{equation}
where: $R$ is 4-dimensional Ricci scalar, $\mathcal{M}$ represents Manifold of spacetime and $g$ is determinant of metric tensor with signature (-,+,+,+). $\mathcal{M}$ can be foliated in $ \mathcal{M}\cong \Sigma \times \mathcal{R}$ thus the action \eqref{A.1.1} can be re-express by the Arnowitt-Deser-Misner action:
\begin{equation}\tag{A.2}
S_{EH}=\frac{c^4}{2\kappa}\int_{\Sigma \times \mathcal{R}} dtd^3xN\sqrt{h}(R+K^{ij}K_{ij}+K^2)
\label{A.1.2}
\end{equation}
where: $h$ is a determinant of metric on one 3-dimensional hypersurface $\Sigma_t$, $N$ is the lapse function which expresses the proper time elapsed between the event belonging in $\Sigma_t$ and $\Sigma_{t+dt}$ (with shift vector denoted by $N^i$ they serve as transformation to obtain every event on the spacetime) and $R$ is now 3-dimensional Ricci scalar \cite{Kiefer:2007ria}. $K^{ij}$ is extrinsic curvature that can be understood as the rate of change of $h_{ij}$ with respect to the global time.\\
\indent By obtaining Hamiltonian density by standard Legendre transformaition, \eqref{A.1.2} takes form:
\begin{equation}\tag{A.3}
S_{EH}=\frac{1}{2\kappa}\int_{\Sigma \times \mathcal{R}} dtd^3x(p^{ij}\dot{h}_{ij}-N\mathcal{H}_{g}-N^i\mathcal{H}_i)
\label{A.1.3}
\end{equation}
where: $p^{ij}$ is conjugate momentum to $h_{ij}$. They Poisson Bracket satisfy:
\begin{align*}
\{p^{ij}(x), h_{kl}(x')\}=\delta^i_k\delta^j_l\delta(x-x') \tag{A.4}
\label{A.1.4}
\end{align*}
What is left is gravitational Hamiltonian constraint $\mathcal{H}_{g}$ and diffeomorphism constraint $\mathcal{H}_i$ \cite{Kiefer:2007ria} that, alongside the addition of Hamiltonian equation of motion, recreates physics. \\
\indent When triads formalism is taken into consideration 3 metric takes form 
$$h_{ij}=e^a_ie^b_j\delta_{ab} $$
and new variables can be introduce (a,b = 1, 2, 3 are space index
and i,j = 1, 2, 3 are internal indices enumering the vectors). Playing the role of $h_{ij}$ is not triad itself but it densitize version:
$$
E^a_i=\sqrt{h}e^a_i
$$
and it canonically conjugate momentum is given by connection:
\begin{equation} \tag{A.5}
A^i_a=\Gamma^i_a+\gamma K^i_a
\label{taga5}
\end{equation}
where $K^i_a$ is externistic curveture 1-form, $\Gamma^i_a$ encodes information about the spin connection. Corresponding to \ref{A.1.4}, Poisson Bracket of these new variables is:
\begin{equation}\tag{A.6}
\{A^i_a(x), E^b_j(x')\}=\kappa\gamma\delta^i_j\delta^b_a\delta(x-x')
\label{A.1.5}
\end{equation}
\indent But changing $(h,p) $ to $(A, E)$ is not enough to obtain the same physics, because the considered system obtains additional degrees of freedom. To fix that, the addition of a new constraint is needed. Thanks to that, \ref{A.1.3} is re-expressed as:
\begin{equation}\tag{A.7}
S_{EH}=\frac{c^4}{2\kappa}\int_{\Sigma \times \mathcal{R}} dtd^3x(E^a_i\dot{A}_{a}^i-N\mathcal{H}_{g}-N^a\mathcal{H}_a-\mathcal{N}^i\mathcal{G}_i)
\label{A.1.6}
\end{equation}
where $\mathcal{N}^i$ is Lagrange multiplier of the additional constraint $G_i$. Such constraint referred to in literature \cite{Kiefer:2007ria} as Gaus Constraint.\\
\indent Now constraints in connection variables takes explicit form:

\begin{align}\tag{A.8.a}
\mathcal{G}_i &= \frac{1}{2\kappa\gamma} \int_\mathcal{V}d^3x  (\partial_a E^a_i +\epsilon_{ijk} A^j_a E^a_k) \\
\mathcal{H}_a &= \frac{1}{\kappa \gamma } \int_\mathcal{V}d^3x F^i_{ab} E^b_i  \tag{A.8.b}\\
\mathcal{H}_g &= \frac{1}{2\kappa}\int_\mathcal{V}d^3x\frac{E^{ai} E^{bj}}{\sqrt{det(h)}}(\epsilon_{ijk}F^k_{ab}-2(1+\gamma^2)K^j_{[a}K^i_{b]}) \tag{A.8.c}
\label{A.1.7}
\end{align}
where:
\begin{equation}\tag{A.9}
H^E=\frac{1}{2\kappa}\int_\mathcal{V}d^3x\epsilon_{ijk}\frac{E^{ai} E^{bj}}{\sqrt{det(h)}}F^k_{ab}
\label{A.1.8}
\end{equation}
\begin{equation} \tag{A.10}
T=\frac{1}{2\kappa}\int_\mathcal{V}d^3x\frac{E^{ai} E^{bj}}{\sqrt{det(h)}}K^j_{[a}K^i_{b]}
\label{A.1.9}
\end{equation}
are Euclidean Hamiltonian constraint and Lorentzian Hamiltonian constraint respective \cite{Ashtekar:2011ni, Thiemann:2007zz}. Integration here is limited to an elementary cell $\mathcal{V}$ , and $F^k_{ab}$ is curvature of connection $A^i_a$ define as: 
$$F^k_{ab}=\partial_a A^k_b-\partial_b A^k_a+\epsilon_{kij}A^i_aA^j_b $$

\section{Deficiency Subspaces}
\label{appB}
Here we recal Corollary from \cite{Reed:1975uy}. If we consider $T$ to be a symmetric operator on a Hilbert space $\mathcal{H}$ then:
\begin{align*}
\mathcal{K}_+=Ker(i-T^{\star}) \\
\mathcal{K}_{-}=Ker(i+T^{\star})
\end{align*}
are deficiency subspaces of $T$. In addition lets define a pair of numbers:
\begin{align*}
n_+(T)=dim(\mathcal{K}_+) \\
n_-(t)=dim(\mathcal{K}_{-})
\end{align*}
which  are called the deficiency indices of $T$. If:
\begin{itemize}
\item[(i)] $n_+(T)=n_-(t)=0$, $T$ is self-adjoint.
\item[(ii)] $n_+(T)=n_-(t)$, $T$ has self-adjoint extensions. There is a one-to-one
correspondence between self-adjoint extensions of $T$ and unitary maps from $\mathcal{K}_+$ onto $\mathcal{K}_{-}$.
\item[(iii)] $n_+(T)\neq n_-(t)$, $T$ has no nontrivial symmetric extensions.
\end{itemize}
\end{appendices} 
\newpage
\bibliographystyle{plain}
\bibliography{magb}

\end{document}